\documentclass[reprint,prd,twocolumn,superscriptaddress,showpacs,nofootinbib,floatfix]{revtex4-1}

\usepackage[dvips]{graphicx}
\usepackage{epsfig,amsmath,amssymb,verbatim,mathrsfs,array,layout,textcomp,latexsym,slashed,graphicx,bbm}

\usepackage[normalem]{ulem}
\usepackage{appendix}

\usepackage{rotating}
\usepackage{hyperref}
\usepackage[usenames,dvipsnames]{color}

\usepackage{soul}

\newcommand{\be}{\begin{equation}}
\newcommand{\ee}{\end{equation}}
\newcommand{\bea}{\begin{eqnarray}}
\newcommand{\eea}{\end{eqnarray}}

\definecolor{gbcolor}{rgb}{.8,.3,.1}
\definecolor{gbcolor2}{rgb}{.8,.1,.7}

\usepackage[utf8]{inputenc}
\usepackage[english]{babel}

\def\beq{\begin{equation}}
\def\eeq{\end{equation}}

\begin{document}


\title{On the Stability of Infinite Derivative Abelian Higgs }

\author{Anish Ghoshal}
\affiliation{Theory Group, Laboratori Nazionale di Frascati-INFN, C.P. 13, 100044, Frascati, Italy}
\affiliation{Dipartimento di Matematica e Fisica, University Roma Tre, Via della Vasca Navale, 84, Rome-00146, Italy}
\author{Anupam Mazumdar}
\affiliation{Van Swinderen Institute, University of Groningen, 9747 AG Groningen, The Netherlands}
\affiliation{Kapteyn Astronomical Institute, University of Groningen, 9700 AV Groningen, The Netherlands}
\author{Nobuchika Okada}
\author{Desmond Villalba}
\affiliation{Department of Physics and Astronomy, University of Alabama, Tuscaloosa, Alabama 35487, USA}

\begin{abstract}
Motivated by the stringy effects by modifying the local kinetic term of an Abelian Higgs field by the Gaussian kinetic term we show that 
the Higgs field does not possess any instability, the Yukawa coupling between the scalar and the fermion, the gauge coupling, and the self interaction
of the Higgs yields exponentially suppressed running at high energies, showing that such class of theory never suffers from 
vacuum instability.  We briefly discuss its implications for the early Universe cosmology.
\end{abstract}

\maketitle

\section{Introduction} 
One of the key aspects of quantum field theory is to remove the ultraviolet (UV) divergences as in the case of quantum electrodynamics. In 
the context of Standard Model (SM), however, this leads to the {\it well-known} hierarchy problem for  the SM Higgs~\cite{Haber:1984rc}. 
The {\it key} observation is that the bosonic  and the fermionic loop corrections to the Higgs mass and the self-interacting term yields an 
opposite effect, which does not cancel exactly, but leads to seventeen-orders of magnitude difference between the electroweak scale 
and the Planck scale. Furthermore, negative running of the self-interacting term for the  SM Higgs gives rise to a metastable vacuum 
roughly at $10^{11}$~GeV, depending on the pole mass of the top-quark at the electroweak scale, see~\cite{Olive:2016xmw}.

Indeed, there are many solutions to this problem including that of supersymmetry at low scales, see~\cite{Haber:1984rc}. Here, we aim to address these problems not
by introducing new states closer to the electroweak scale, but by invoking inherently another new scale  motivated by string field theory by modifying the kinetic operator of the field with infinite derivatives~\cite{sft1,sft2,sft3,padic1,padic2,padic3,Frampton-padic,Tseytlin:1995uq,marc,Siegel:2003vt}. Such infinite derivatives indeed  soften the UV behaviour, if they are captured by an {\it exponential of an entire function}~\cite{Biswas:2014yia}. First of all, exponential of an entire function does not introduce new poles in the propagator, therefore no new degrees of freedom arises
in the spectrum~\cite{Biswas:2005qr, Biswas:2011ar}. The propagator is exponentially suppressed in the UV, which helps softening the UV quantum behaviour.

In fact, it has been recently noticed in the context of gravity that an infinite derivative gravity, the gravitational interaction becomes weak at short distances and small time scales, such that the Big bang Singularity is now replaced by the Big bouncing, non-singular, cosmology, see~\cite{Biswas:2005qr,Biswas:2011ar}. Furthermore, there is also no blackhole type singularity, because gravitational interaction can be weakened at a macroscopic scale of Schwarzschild's radius~\cite{Koshelev:2017bxd}, while recovering the Newtonian $1/r$-fall at large distances from the source~\cite{Biswas:2011ar,Edholm:2016hbt}. Such theories also behave better in the quantum aspect, see~\cite{Tomboulis,Modesto,Talaganis:2014ida}, where infinite derivative gravity becomes finite in the UV. Similarly, modifying the kinetic term for the scalar, fermionic, and gauge field theory~\cite{Biswas:2014yia} have been studied in the past, see \cite{Wataghin,Efimov:1967pjn,Efimov-unitarity,moffat-prd,moffat-qft,efimov-qed1,efimov-qed2,Moffat:2011an}. 

The aim of this paper is to show that in the case of an Abelian Higgs model
the presence  of {\it infinite derivatives} in the kinetic term yields all the $\beta$-functions to become
exponentially suppressed in the UV. Indeed, this is one promising way to ameliorate the stability problem of the SM Higgs, because 
both Abelian and SM Higgs share similar interactions and properties. Towards the end we will comment on non-Abelian structure as well.

\section{A simple toy model}
In order to illustrate this, let us first consider a simple scalar toy model with $\phi$ being a real scalar and $\psi$ being a Dirac fermion, 
the Lagrangian is \cite{Biswas:2014yia},
\begin{eqnarray}
 \mathcal{L} &=& - \frac{1}{2} \phi e^{{\Box + m_{\phi}^{2}\over M^2}} \left(\Box + m_{\phi}^{2}
  \right)\phi 
 + \bar{\psi} e^{{\Box + m_{\psi}^{2}\over M^2}}   \left(i \gamma ^\mu \partial_{\mu} -m_\psi \right)  \psi  \nonumber\\
&&   -\frac{\lambda}{4 !} \phi^4 - y \phi \bar\psi \psi. 
\end{eqnarray}
where $\Box= \eta_{\mu\nu}\partial^{\mu}\partial^{\nu}$ $(\mu, \nu=0,1,2,3$) with the convention of the metric signature $(+,-,-,-)$, 
  $m_{\phi}$ and $m_{\psi}$ are the masses of the scalar and the Dirac fermion, respectively, 
  and $M$ is a scale of the non-locality which is taken to be below Planck scale. 
In our non-local field theory, only the kinetic terms are modified with the non-local scale $M$, 
 while the scalar self-interaction and the Yukawa interaction are the standard one. 
The theory is reduced into the standard local field theory in the limit of $M \to \infty$.
In the Euclidean space $(p ^0 \rightarrow ip_E^0 )$ the scalar and fermion propagators are given by~\cite{Biswas:2014yia}:
\begin{equation}
\Pi_\phi(p_E) = - {ie^{-{p_E^2 + m_\phi^2\over M^2}}\over p_E^2 + m_\phi^2}, \; 
\Pi_\psi(p_E) = -{i \slashed{p}_E e^{-{p_E^2 + m_\psi^2\over M^2}}\over p_E^2 + m_\psi^2}\,.
\end{equation}
Note that the non-local extension of the theory leads to the exponential suppression of the propagators for $p_E^2 > M^2$, 
and this fact indicates that quantum corrections will be frozen at energies higher than $M$. 
Since the mass terms for the scalar and fermion fields are irrelevant to our discussion, 
we drop them in the following analysis. 

Let us now consider renormalization group (RG) evolutions of the couplings $\lambda$ and $y$. 
The one-loop $\beta$-function for $\lambda$ is of the form, 
\begin{eqnarray} \label{E1}
\mu \frac{d \lambda}{d \mu} =
\beta_{\lambda}^{(1)} 
= f_{\lambda\lambda} \lambda^2 + f_y y^4 +  f_{\lambda y} \lambda y^2 ,
\end{eqnarray}
where $\mu$ is the renormalization scale, $f_{\lambda \lambda} \lambda^2$ and  $f_y y^4$ are due to the quantum corrections 
  to the self-interaction vertex through scalar and fermion loop diagrams, respectively, 
  while $f_{\lambda y}y^2 \lambda$ is from the wave function renormalization of $\phi$ 
  (anomalous dimension).
In the standard quantum field theory (QFT), $f_{\lambda \lambda} \lambda ^2$ and $f_y y^4$ 
  can be extracted from the coefficients of logarithmic divergences 
  of the 4-point 1-PI function in the loop-integrals. 
Although the propagators are modified in our present scenario, 
  we adopt this standard procedure, in order to recover the standard QFT results in the limit of $M \to \infty$. 
Setting all external momenta to be zero, the 4-point 1-PI function is given by 
\bea \label{Vc1}
 \Gamma_{4} &=&
-\frac{3}{2} \lambda^2 \int \frac{d^4 k_E}{(2 \pi)^4} 
 \left( \Pi_\phi(k_E) \right)^2    \nonumber \\
&&
-6 y^4 \int  \frac{d^4 k_E}{(2 \pi)^4} 
\,{\rm tr} 
\left[ \left( \Pi_\psi(k_E) \right) ^4
\right] \nonumber \\
&=&
 \int^{\Lambda^2} d k_E^2 
 \left( 
 \frac{3 \lambda^2}{32 \pi^2}
 \frac{e^{-2 \frac{k_E^2}{M^2}}}{k_E^2} 
- \frac{24 y^4 }{16 \pi^2}
 \frac{e^{-4 \frac{k_E^2}{M^2}}}{k_E^2} 
\right) , 
\eea
where we have introduced a cutoff scale $\Lambda$ for the loop-integral 
  in the Pauli-Villars regularization scheme.\footnote{Pauli-Villars scheme is to show the connection between conventional QFT and our non-local case, 
based on our criterion that the conventional QFT result must be recovered in the limit of M $\rightarrow \infty$. However, loop-integral between $\mu$ and $\mu + \delta \mu$ can also be considered to 
perform the integrals since the integrals do not diverge due to the exponential softening.} 
According to the standard QFT procedure, we extract the $\beta$-function 
  from $\Gamma_4$ by taking $\partial/\partial \log \Lambda$ and replacing the cutoff into 
  the renormalization scale $\mu$: 
\bea 
  f_{\lambda\lambda} \lambda^2 + f_y y^4 &=& 
  \Lambda \frac{\partial}{\partial \Lambda} \Gamma_4 \Bigg|_{\Lambda \to \mu}  \nonumber \\
  &=& \frac{3 \lambda^2}{16\pi^2}e^{-2 \frac{\mu^2}{M^2}} 
 -\frac{48 y^4}{16\pi^2}e^{-4 \frac{\mu^2}{M^2}} .
\eea
Note that the $\beta$-function is vanishing for $\mu > M$, 
 while it retains the usual form upon taking $M \to \infty$ as we required. 
The power of the exponent $\left( {\rm exp} \left[- \mu^2/M^2 \right] \right)^n$ is determined 
  by the number of propagators running in the loop diagrams. 

With an external momentum $p_E$ in the Euclidean space, 
  the 2-point 1-PI function is given by  
\bea
\Gamma_{2} = y^2 
\int \frac{d^4 k_E}{(2 \pi)^4} \,{\rm tr} 
\left[ \Pi_\psi\left(k_E \right) \, \Pi_\psi\left(k_E + p_E \right)
\right] 
\label{Gamma_2}
\eea  
Since the non-local extension involves quit a non-trivial $p_E$ dependence 
  for the loop-integral, the wave function renormalization term is more cumbersome to calculate,  
  though we expect in the high energy limit ($\mu^2 \gg p_E^2$), 
  the effect of the external momentum on the correction is negligible and we find 
\be \label{Ee2}
 f_{\lambda y} \lambda y^2 = 4 \lambda \gamma_\phi \approx
   \frac{8\lambda y^2}{16 \pi^{2}}e^{-2 \frac{\mu^2}{M^2}}, 
\ee
where $\gamma$ is the anomalous dimension for the scalar filed $\phi$.  
For reader's convenience, we present detailed calculations of the anomalous dimension in Appendix A.
Since the wave function renormalization is also exponentially suppressed,
the field itself is completely frozen at energies higher than $M$.

Now we have found the $\beta$-function for $\mu > M$ is modified and approximately given by 
\bea
 \beta_\lambda^{(1)} \approx  \frac{1}{16\pi^2} 
 \left( 3 \lambda^2 +8 \lambda y^2 \right) e^{-2 \frac{\mu^2}{M^2}}.  
\eea
Very interestingly, the $\beta$-function is vanishing towards high energies, and therefore  
   the theory approaches a conformal field theory in the UV limit. 
In the view point of the vacuum stability, the dangerous correction factor $f_y$, 
  which ordinarily drives the running $\lambda(\mu)$ to become negative, 
  have become tamed and ultimately irrelevant in the UV limit. 
Hence,  the non-local modifications to the quantum corrections provides a pathway to remedy
  the Higgs vacuum instability problem in the Standard Model!

Similarly, for the Yukawa interaction, we calculate the $\beta$-function 
  for the Yukawa coupling ($y$), 
\bea 
\mu \frac{d y}{d \mu} =
 \beta^{(1)}_y = \left( f_{w}   + f_{v} \right) y^3, 
\eea 
where $f_{w} y^3$ is from the wave function renormalizations for the scalar and the fermion, 
 while $f_{v} y^3 $ is from one-loop corrections to the 3-point 1-PI function. 
Setting all external momenta to be zero, the 3-point 1-PI function is given by 
\bea
 \Gamma_3 &=& i y^3  \int \frac{d^4 k_E}{(2 \pi)^4} 
   \Pi_\phi(k_E) \,  \left(\Pi_\psi(k_E) \right)^2  \nonumber \\
&=& \frac{y^3}{16 \pi^2} 
\int^{\Lambda^2} d k_E^2   \frac{e^{-3 \frac{k_E^2}{M^2}}}{k_E^2}  ,
\eea 
so that 
\bea 
f_v y^3 = \Lambda \frac{\partial}{\partial \Lambda} \Gamma_3 \Big|_{\Lambda \to \mu} 
  = \frac{2}{16\pi^2} y^3 e^{-3 \frac{\mu^2}{M^2}} .
\eea
The anomalous dimension for the fermion is calculated in Appendix B, and we obtain
\bea 
  f_w y^3 &=& y \left( \gamma_\phi + 2 \gamma_\psi \right) \nonumber \\
&\approx& 
  \frac{y^3}{8\pi^2} y^3 e^{-2 \frac{\mu^2}{M^2}}  +
  \frac{y^3}{32\pi^2} y^3 e^{-2 \frac{\mu^2}{M^2}}  \nonumber \\
&=&   \frac{3}{16\pi^2} y^3 e^{-2 \frac{\mu^2}{M^2}}  . 
\eea
Thus, we have found the $\beta$-function for $\mu > M$ is modified and approximately given by
\bea 
 \beta^{(1)}_y  &\approx&  \frac{3}{16\pi^2} y^3 e^{-2 \frac{\mu^2}{M^2}}  + \frac{2}{16\pi^2} y^3 e^{-3 \frac{\mu^2}{M^2}} \nonumber\\
&\approx & \frac{3}{16\pi^2} y^3 e^{-2 \frac{\mu^2}{M^2}} 
\eea  
The Yukawa coupling running also behaves in the similar manner to the running of $\lambda(\mu)$, 
  where its $\beta$-function vanishes toward high energies.

\begin{figure}[t] 
\centering
\includegraphics[height=5cm, width=7cm]{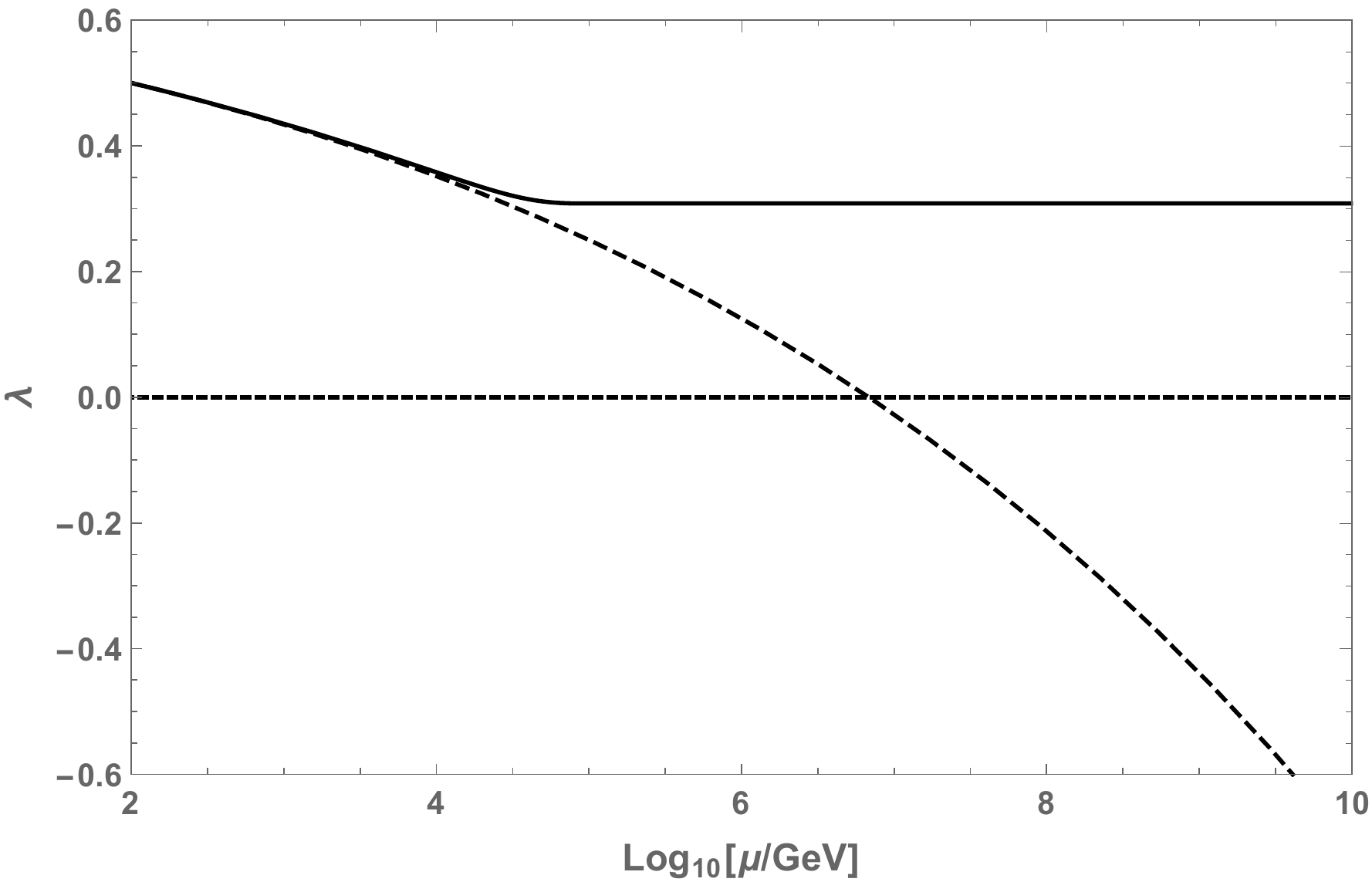} \hspace{1cm}
\caption{
The scalar self-coupling running is shown in  solid (dashed) black lines for the non-local (local) behavior. 
Here, we have taken initial values $\lambda=0.5$ and $y=0.6$ at $\mu=100$ GeV, and the non-local scale to be $M=10^5$ GeV. 
}
\label{f2}
\end{figure}

In Figure \ref{f2}, we show the RG evolution of the scalar self-coupling 
in the non-local theory (solid  line), along with the one in the standard theory (dashed line). 
Here, we have taken $\lambda=0.5$ and $y=0.6$ at $\mu=100$ GeV, and the non-local scale to be $M=10^5$ GeV. 
As we expect, the scalar self-coupling stops its evolution around $\mu=M=10^5$ GeV, 
  while the running coupling in the standard theory becomes negative 
  by the negative contribution form $f_y$.

\begin{figure}[t] 
\centering
\includegraphics[height=5cm, width=7cm]{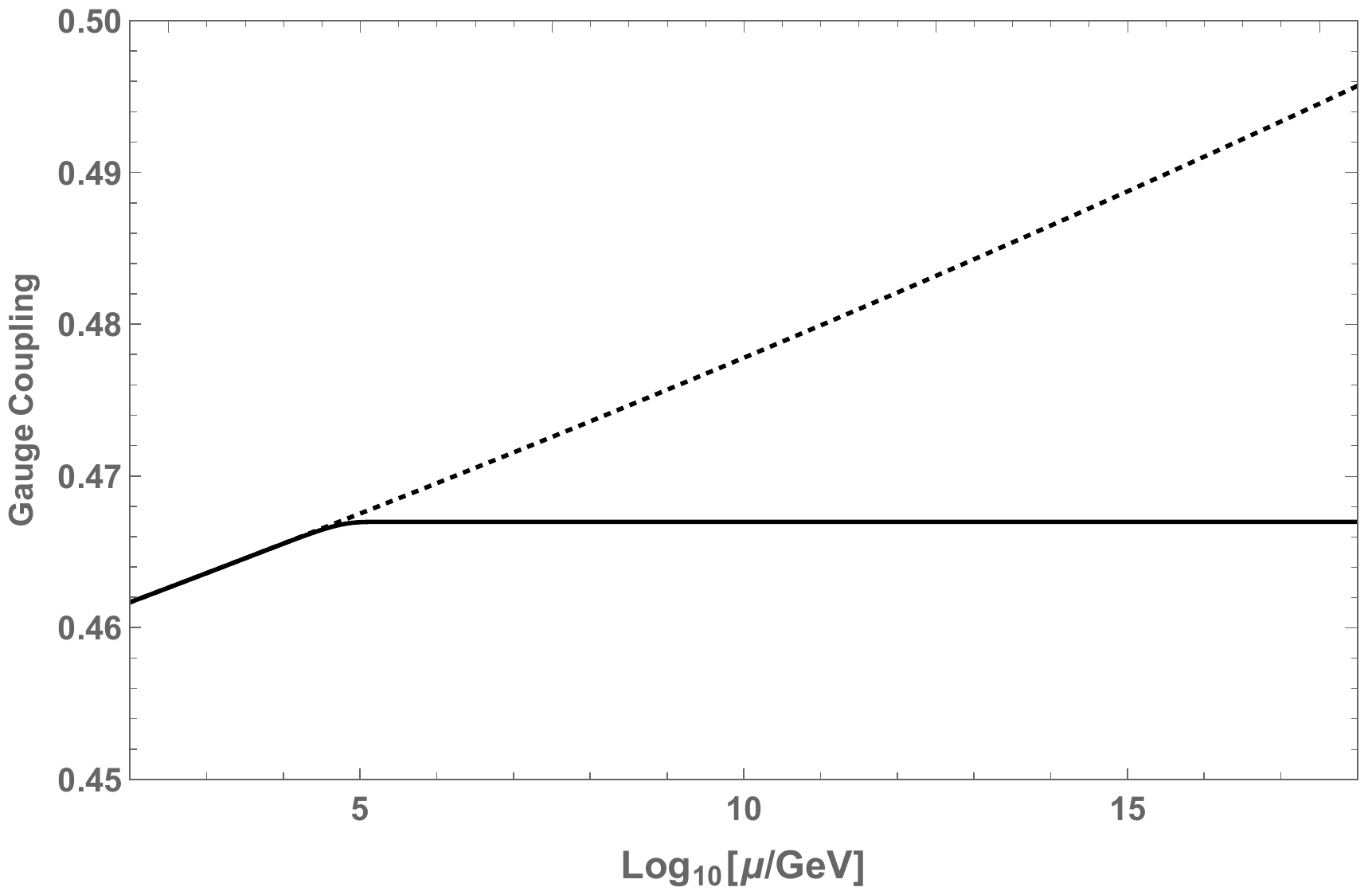} \hspace{1cm}
\caption{
The running U(1) gauge coupling shown in solid (dashed) black lines for the non-local (local) theories. 
Here, we have taken $M=10^5$ GeV.}
\label{f5} 
\end{figure}

\section{Non-local QED}
Next we study the gauge coupling running in the non-local QED. 
The incorporation of a gauge interaction is much the same as in the local QED case. 
Now due to the non-local nature of the theory the field strength is modified along with the covariant derivative expression. 
Consider the following gauge invariant Lagrangian for the non-local QED with a Dirac fermion \cite{Biswas:2014yia},
\be \label{L1}
\mathcal{L} = -\frac{1}{4} F ^{\mu\nu} e ^{\frac{\Box}{M ^2}} F _{\mu\nu} + i\bar{\psi} e ^{\frac{\nabla^2}{M ^2}} \gamma ^{\mu} D_{\mu} \psi 
  + {\rm H.c.} ,
\ee
where $D_{\mu}=\partial_{\mu} + i g A_{\mu}$  is the covariant derivative and $\nabla^2=D_{\mu} D^{\mu}$. 
The covariant form in the exponential found in the fermion term is necessary, along with the hermitian conjugate of it, 
  to ensure hermiticity of the Lagrangian.

Although Abelian gauge theories are asymptotically non-free in the standard QFT, 
  their non-local extension results in a very interesting behavior. 
As in the standard QFT, we extract the $\beta$-function of the gauge coupling 
  from the anomalous dimension of the gauge field as $\beta_g = \gamma_A$, which is calculated in Appendix C.

As mentioned previously, calculations of any wave function renormalizations is complicated. 
Borrowing our approach in determining the $f_{\lambda y}$ term above and assuming  $\mu^2 \gg p_E^2$, 
 we obtain the following expression for the RG equation of the gauge coupling (see Appendix C):
\bea \label{f1}
\mu\frac{dg}{d\mu}=\frac{1}{16\pi^{2}} \left(\frac{4}{3}\right)  g^3 e^{-2 \frac{\mu^2}{M^2}}. 
\eea
Clearly, the standard result retains in the limit of $M \to \infty$.
Similarly to the RG evolutions of the scalar self-coupling and the Yukawa coupling, 
  the influence of the non-local effects has essentially rendered a would-be asymptotic non-free theory UV finite or "UV-insensitive". 
From Eq.~(\ref{f1}), it is evident that once the running of the couplings passes beyond the non-local scale $M$, 
  the $\beta$-function rapidly approaches zero, resulting in a constant value of coupling as seen in Figure \ref{f5}.

The advantage of a Gaussian kinetic operator is that the Abelian Higgs becomes scale free beyond energy scale, $M$. The theory becomes
conformal, and $M$ signifies the UV-fixed point. Conformal theory has many classical and quantum virtues, although the theory becomes 
trivial beyond $M$ here, it is pertinent to think about how to generate the value of $M$. At this point, we merely speculate this scale beyond the SM and 
Planck scale, but its origin might arise from the scale of compactification from higher dimensions, or from the VEV of the dilaton, for instance in string theory~\cite{marc,Siegel:2003vt}.

\section{Applications to cosmology}
One particular application for the scale free theory comes in the context of cosmological inflation, for a review see~\cite{Mazumdar:2010sa}. In this
respect the value of $M$ will play a significant role for breaking the scale invariance, as well as creating the density fluctuations observed 
in the cosmic microwave background radiation~\cite{Ade:2015xua}.  

The freezing of the self-interaction term in the UV leads to a scale free potential beyond the energy scale of $M$. The field $\phi$ would 
be the ideal candidate for inflaton  driving potential dominated inflation, since the fermion coupling also freezes, the radiative correction 
to $\phi$ from fermion-loop will not spoil the flatness of the potential all the way beyond $M< M_p\simeq 2.4\times 10^{18}$ GeV, and even beyond $M_p$.
The potential maintains a perfect shift-symmetry, therefore $\phi$ can take any VEVs. In this respect realising super-Planckian VEV during 
inflation will not be forbidden, albeit there are quantum gravity corrections which are expected to be tiny, as long as 
the energy density of the inflationary vacuum is still below the Planckian energy density.

In the simplest scenario with $\lambda \phi^4$ potential the inflationary predictions are already ruled out by the current data 
from Planck~\cite{Ade:2015xua}, nevertheless a slight modification in the setup such as the SM Higgs inflation~\cite{Berzukov}, might be a simple
way to realise a falsifiable model of inflation.  Indeed, the scale of non-locality will play a significant role here, which we will explore in future studies.  
Nevertheless, we can make some intriguing comments here already. Typically, in this setup we would require minimal kinetic term for Higgs, which is  
coupled to the Ricci scalar via a non-minimal coupling. In order to explain the amplitude of the observed density perturbations in the cosmic microwave background radiation (CMBR), we would expect the non-mininal coupling to be large of the order of ${\cal O}(4000)$, in order to sufficiently flatten the potential~\cite{Berzukov}. 
  In our case, though, this non-minimal coupling can be made even order ${\cal O}(1)$, the flattening of the potential will essentially comes from the scale free theory, and 
the end of inflation will arise from the RG equation, which modifies the potential close to the scale of non-locality $M$. Indeed, all the details have to be worked out 
carefully to show that this mechanism can work properly for a viable model of inflation. 

A point to contemplate in this regard is that besides the initial conditions which are to be chosen in order to evolve the differential equations
to a point where the observables are measured or the Universe behave as it does today, we have introduced a degree of freedom (at least from mathematical point 
of view) choosing which suitably may give us the correct Universe even with starting off from an initial condition otherwise considered incompatible.
  
There is another intriguing feature for thermal history of the Universe. Since, all the interactions freeze above $M$, the scattering rate, $\Gamma$, between 
species, i.e. scalar, fermion and Abelian gauge field, will not be able to cope with the Hubble expansion rate, $H$, of the Universe, i.e. $\Gamma \ll H$.  Indeed, the expansion rate of the Universe would still be dominated by the dynamics of the scalar field, however, the Universe would be effectively {\it cold} for energy scale $\geq M$. The value of $M$ would effectively determine the dynamics of reheating of the Universe, once the {\it big-freeze} in the interactions cease. This limits the reheat temperature of the Universe $ T_{rh} \leq M$ for all practical purposes.

\section{Conclusions}
We have shown that within the Abelian Higgs model, a Gaussian kinetic operator, which introduces no new poles other than the original theory gives rise to 
a scale-invariant or a conformal theory in the UV beyond the scale of non-locality. The presence of non-locality arises in the interactions when the vertex operators gain
exponential factors in the momentum space, smearing out the vertices. Infinite derivatives are required precisely to make the theory {\it ghost free} all the way from IR to UV. In this regard what we have shown is that the beta functions for the Abelian Higgs and the fermion exponentially decreases in the UV beyond the scale of non-locality, essentially making the theory dynamically stable beyond $M$. Indeed, the choice of $M$ here is arbitrary. However, it is indicative that in the non-Abelian Higgs, such a new scale if appears around $10^{10}-10^{11}$~GeV would definitely yield a stable SM Higgs without invoking any new symmetries. Our results open a new way to model the stability issue concerning any scalar field theory, a detailed RG equations of non-Abelian Higgs will be presented in a separate publication.

Finally, we wish to  make a comment on introducing non-locality for the non-Abelian Higgs and SM fermions. Despite the $SU(2)_L$ structure
making the loop computations a bit more complicated, the final outcomes are very similar to that of the Abelian Higgs in presence of non-locality. The 
Gaussian kinetic term will always pave the exponential suppression in the propagator in the UV, while enhancing the vertex operators. It is easy to show that the self-interacting non-local terms from the non-Abelian generators (arising due to covariant derivatives) only contribute as higher and higher dimensional operators thus highly suppressed by powers of M. 
The UV behaviour of the Higgs will be very similar to what we have discussed in the Abelian case, including the leading order behaviour of the beta-functions for the SM Higgs and the fermions.

\textit{Acknowledgments:} 
Authors would like to thank Enrico Nardi, Davide Meloni, Arindam Chatterjee and Alessandro Strumia for discussions. 
The work of A.G. was supported by ISI-Kolkata, Roma Tre and LNF-INFN facilities.
The work of N.O. is supported in part by the United States Department of Energy (DE-SC0012447).\\

\begin{center}
\noindent{\large \bf Appendix}
\end{center}
\appendix
Our methodology for calculating the wave function renormalization follows from the same procedure 
  of calculating quantum corrections in conventional quantum field theories. 
As all quantum corrections are switched off at energies beyond scale $M$ in non-local theories, 
  strictly speaking there are no corrections in this UV regime. 
However, our guiding  philosophy is that as the non-local scale is pushed towards the limit  of $M\to \infty$, 
  we need to recover all the results in the conventional quantum field theories. 

\section{Scalar Wave Function Renormalization}
The 2-point 1-PI function in Eq.~(\ref{Gamma_2}) is explicitly given by
\bea
\Gamma_{2} &=& y^2 \int \frac{d^4 k_E}{(2 \pi)^4} \,{\rm tr} 
\left[ \Pi_\psi(k_E) \, \Pi_\psi(k_E+p_E)  \right] \nonumber \\
&=& 
y^2 \int \frac{d^4 k_E}{(2 \pi)^4} \,{\rm tr} 
\left[ \Pi_\psi(k_E - p_E/2) \; \Pi_\psi(k_E+p_E/2)
\right] \nonumber \\
&=& 
4 y^2\int \frac{d^4 k_E}{(2 \pi)^4} \frac{(k_E - \frac{p_E}{2})\cdot(k_E+\frac{p_E}{2})}{(k_E-\frac{p_E}{2})^2(k_E+\frac{p_E}{2})^2} \\ \nonumber 
&&\times \, e^{-\beta(k_E-\frac{p_E}{2})^2}e^{-\beta(k_E+\frac{p_E}{2})^2}, 
\eea
where $\beta=1/M^2$.  
Using the Schwinger parameters, $\alpha_1$ and $\alpha_2$, and completing the square in the exponent yields
\bea \label{Vc1}
 \Gamma_{2} &=&
4 y^2 \int \frac{d k_E^4}{(2 \pi)^4}  \int_0^\infty d\alpha_1 d\alpha_2  
\left(k_E-\frac{p_E}{2}\right)\cdot\left(k_E+\frac{p_E}{2}\right)   \nonumber \\
&&\times
 e^{-\left((\alpha_1+\alpha_2+2\beta)^{1/2}k_E+\frac{\alpha_1-\alpha_2}{(\alpha_1+\alpha_2+2\beta)^{1/2}}\frac{p_E}{2}\right)^2} 
  \nonumber \\
&&\times 
e^{-\left(\alpha_1+\alpha_2+2\beta-\left(\frac{\alpha_1-\alpha_2}{(\alpha_1+\alpha_2+2\beta)^{1/2}}\right)^2\right)\frac{p_E^2}{4}}.
\eea
Introducing new parameters defined as $\alpha_1+\alpha_2=s$ and $\alpha=\alpha_1/s$, 
  and shifting the momentum by $ k_E \to k_E- \frac{s(2\alpha-1)}{s+2\beta}\frac{p_E}{2}$, 
  we express $\Gamma_{2}$ as 
\bea \label{Vc1}
\nonumber
 \Gamma_{2} &=&
4 y^2 \int \frac{d k_E^4}{(2 \pi)^4} 
\int_{0}^{1} d\alpha \int_{0}^{\infty} s ds   \nonumber\\
&&
\times
 \left( k_E^2 - \left(1- \left(\frac{s (2\alpha-1)}{ (s+2\beta)}\right)^2 \right)\frac{p_E^2}{4}\right) \nonumber \\
 &&
\times \, e^{-\left[(s+2\beta) k_E^2
+\left(s+2\beta-\left(\frac{s(2\alpha-1)}{(s+2\beta)^{1/2}}\right)^2\right)\frac{p_E^2}{4}\right]} .
\eea
We extract the wave function renormalization factor as a coefficient of the external momentum $(-p_E^2)$ by 
\bea
&& \frac{\partial}{\partial (-p_E^2)} \Gamma_2 \Bigg|_{p_E^2=0}=
\frac{y^2}{16\pi^2}\int^{\Lambda^2} k_E^2 d k_E^2\int_{0}^{1} d\alpha \int_{0}^{\infty} s ds \nonumber \\
&& \times \, e^{-(s+2\beta) k_E^2} \nonumber\\
&&
\times \left(k_E^2\left(s+2\beta-\left(\frac{s(2\alpha-1)}{(s+2\beta)^{1/2}}\right)^2\right) +1-\left(\frac{s(2\alpha-1)}{s+2\beta}\right)^2\right) \nonumber\\
&&\approx 
 \frac{y^2}{16 \pi^2}\int^{\Lambda^2}k_E^2 d k_E^2 \, \left\lbrace
\frac{3+2\beta k_E^2}{k_E^4} \right.  \nonumber \\
&&
\left.-\frac{1}{3}\left[\frac{2-2\beta k_E^2+4\beta^2 k_E^4}{k_E^4}+\frac{1 -4\beta k_E^2-4\beta^2 k_E^4}{k_E^4}\right]\right\rbrace e^{-2\beta k_E^2} 
\nonumber \\
&&
=\frac{2 y^2}{16 \pi^2}\int^{\Lambda^2} d k_E^2 \,\frac{e^{-2\beta k_E^2}}{k_E^2}. 
\eea
Here, the approximation indicates that terms which rapidly approach zero for $k_E^2 > M^2$ 
  are excluded for simplicity. 
Applying the standard QFT procedure, we obtain the anomalous dimension as 
\bea 
   \gamma_\phi &=& \frac{1}{2}  \Lambda \frac{\partial}{\partial \Lambda} 
   \left[ \frac{\partial}{\partial (-p_E^2)} \Gamma_2 \Bigg|_{p_E^2=0} \right] \Big|_{\Lambda \to \mu} \nonumber\\ 
&   = &
   \frac{y^2}{8 \pi^2}e^{-2\beta\mu^2}.
\eea

\section{Fermion Wave Function Renormalization}
Employing similar techniques and Schwinger parameterizations as outlined for the scalar field, we calculate the fermion wave function renormalization. The 1-loop calculation is given by 
\bea \label{Vc1}
\Gamma_{2} &=& - y^2 \int \frac{d^4 k_E}{(2 \pi)^4} 
  \Pi_\psi(k_E+p_E/2) \, \Pi_\phi(k_E-p_E/2)   \nonumber \\
&=&
- y^2\int \frac{d^4 k_E}{(2 \pi)^4} \frac{\not{\!k_E}+\frac{\not{p_E}}{2}}{(k_E-\frac{p_E}{2})^2(k_E+\frac{p_E}{2})^2} \nonumber\\
&&\times  e^{-\beta(k_E-\frac{p_E}{2})^2}e^{-\beta(k_E+\frac{p_E}{2})^2}
\\ 
&=&
- y^2\int \frac{d^4 k_E}{(2 \pi)^4} 
\int d\alpha_1 d\alpha_2 \left( \not{\!k_E}+\frac{\not{p_E}}{2}\right) \nonumber \\
&&
\times e^{-\left((\alpha_1+\alpha_2+2\beta)^{1/2}k+\frac{\alpha_1-\alpha_2}{(\alpha_1+\alpha_2+2\beta)^{1/2}}\frac{p_E}{2}\right)^2} 
 \nonumber \\
&&
\times e^{-\left(\alpha_1+\alpha_2+2\beta-\left(\frac{\alpha_1-\alpha_2}{(\alpha_1+\alpha_2+2\beta)^{1/2}}\right)^2\right)\frac{p_E^2}{4}}.
\eea
As in the scalar case, redefining the Schwinger parameters and shifting the momentum, we express 
\bea \label{Vc1}
\nonumber  \Gamma_{2} &=&
- y^2\int \frac{d^4 k_E}{(2 \pi)^4}  \int_{0}^{1} d\alpha \int_{0}^{\infty} s ds  \nonumber \\
&\times& \left( \not{\! k_E} + \left(1-\frac{s(2\alpha-1)}{s+2\beta}\right)\frac{\not{p_E}}{2}\right) \\ \nonumber
&\times& e^{-\left[(s+2\beta) k_E^2
+\left(s+2\beta-\left(\frac{s(2\alpha-1)}{(s+2\beta)^{1/2}}\right)^2\right)\frac{p_E^2}{4}\right]}.
\eea
We now extract the wave function renormalization factor as a coefficient of the external momentum $\not{p}_{E}$ by 
\bea
\frac{\partial}{\partial \not{p_E}} \Gamma_{2} \Big|_{p_E=0}&& =
\frac{y^2}{16 \pi^2}\left(\frac{1}{2}\right) \int^{\Lambda^2} k_E^2 d k_E^2 \int_{0}^{1} d\alpha \int_{0}^{\infty} s ds  \nonumber \\
&&\times \, \left(1-\frac{s(2\alpha-1)}{s+2\beta}\right)e^{-\left(s+2\beta\right) k_E^2} \nonumber \\
&&=
\frac{y^2}{16 \pi^2}\left(\frac{1}{2}\right)\int^{\Lambda^2} d k_E^2 \, \frac{e^{-2\beta k_E^2}}{k_E^2}. 
\eea
Hence, we obtain the anomalous dimension for the fermion as 
\bea 
   \gamma_\psi &=& \frac{1}{2}  \Lambda \frac{\partial}{\partial \Lambda} 
   \left[ 
  \frac{\partial}{\partial \not{p_E}} \Gamma_{2} \Big|_{p_E=0} \right] \Big|_{\Lambda \to \mu}  \nonumber \\
&=& 
   \frac{y^2}{32 \pi^2}e^{-2\beta\mu^2}. 
\eea

\section{Gauge Wave Function Renormalization}
The $\beta$-function in Eq.~(\ref{f1}) originates from the fermion loop contribution to 
  the wave function renormalization of the gauge field. 
Employing the similar techniques in Appendices A and B, we express the 2-point 1-PI function of the gauge field as 
\bea \label{Vc2}
\Gamma_{2} &=&
g^2 \int \frac{d^4 k_E}{(2 \pi)^4} \int_{0}^{1} d\alpha \int_{0}^{\infty} s ds \: \left(g_{\rho \nu}g_{\sigma \mu}-g_{\mu \nu}g_{\rho \sigma} \right. \nonumber \\ 
&+&
\left. g_{\rho \mu}g_{\sigma \nu}\right)  
\left( k_E-\left(1+\frac{s(2\alpha-1)}{s+2\beta}\right)\frac{p_E}{2}\right)^{\rho}  \nonumber \\
&\times& \left( k_E+\left(1-\frac{s(2\alpha-1)}{s+2\beta}\right)\frac{p_E}{2}\right)^{\sigma} \nonumber \\
&\times& e^{-(s+2\beta) k_E^2
-\left(s+2\beta-\left(\frac{s(2\alpha-1)}{(s+2\beta)^{1/2}}\right)^2\right)\frac{p_E^2}{4}}.
\eea
Dropping all terms odd in loop momentum, 
  and using $\int d^{4}k k^{\mu}k^{\nu}f(k^2)=\frac{1}{4}\int d^{4}k k^{2} g^{\mu \nu}f(k^2)$, 
  we simplify the 2-point function as 
\bea
\nonumber  \Gamma_{2} &=&
\frac{g^2}{16\pi^2}\int^{\Lambda^2} k_E^2d k_E^2\int_{0}^{1} d\alpha \int_{0}^{\infty} s ds \: \left(-2 k_E^2g_{\mu \nu} -4\phantom{\frac{1}{1}}\right.  \\ \nonumber
&\times& \left. \left(1+\frac{s(2\alpha-1)}{s+2\beta}\right)\left(1-\frac{s(2\alpha-1)}{s+2\beta}\right)\left(\frac{p_{E\mu}p_{E\nu}}{2} - \frac{p_E^2g_{\mu\nu}}{4}\right)\right)\\ \nonumber
&\times& e^{-(s+2\beta) k_E^2
-\left(s+2\beta-\left(\frac{s(2\alpha-1)}{(s+2\beta)^{1/2}}\right)^2\right)\frac{p_E^2}{4}}. \\
\eea
To easily find the wave function renormalization, we focus on the coefficient of terms only proportional to $p_{E \mu}p_{E \nu}$, 
 which gives 
\bea
&&\nonumber \Gamma_{2} \supset 
- \frac{2 g^2}{16\pi^2}\int^{\Lambda^2} k_E^2 d k_E^2\int_{0}^{1} d\alpha \int_{0}^{\infty} s ds \nonumber \\
&&\times \left(1-\frac{s^2(2\alpha-1)^2}{(s+2\beta)^2}\right) e^{-(s+2\beta) k_E^2} \times \left( p_{E\mu}p_{E\nu} \right) \nonumber \\
&&\approx - \frac{g^2}{8 \pi^2} \left( \frac{4}{3} \right) 
 \left[\int^{\Lambda^2} dk_E^2   \frac{e^{-2\beta k_E^2}}{k_E^2} \right] \left( p_{E\mu}p_{E\nu} \right)
\eea
 From this formula, we read off the $\beta$-function in Eq.~(\ref{f1}).


\bibliographystyle{apsrev4-1}

\end{document}